\documentclass[runningheads]{llncs}
\usepackage{amsmath}
\usepackage{amssymb}
\usepackage{mathtools}
\usepackage{booktabs}
\usepackage{microtype}
\usepackage{graphicx}
\usepackage{subcaption}
\usepackage{algorithm}
\usepackage{algorithmic}
\usepackage{xcolor}

\usepackage[T1]{fontenc}
\usepackage{graphicx}
\begin{document}
\title{Spikewhisper: Temporal Spike Backdoor Attacks on Federated Neuromorphic Learning over Low-power Devices}
%
%
%
\author{Hanqing Fu\inst{1} \and
Gaolei Li\inst{1} \and
Jun Wu\inst{1} \and Jianhua Li\inst{1} \and Xi Lin\inst{1} \and Kai Zhou\inst{2} \and Yuchen Liu\inst{3}}
\authorrunning{H. Fu et al.}
\titlerunning{Spikewhisper: Temporal Spike Backdoor Attacks on FedNL}
%
\institute{Shanghai Jiao Tong University \email{\{fuhanqing,gaolei\_li,junwuhn,lijh888,linxi234\}@sjtu.edu.cn}
\and Hong Kong Polytechnic University 
\\ \email{kaizhou@polyu.edu.hk}\\
\and North Carolina State University 
\\ \email{yuchen.liu@ncsu.edu}
\\
}

\maketitle  
%
\begin{abstract}
Federated neuromorphic learning (FedNL) leverages event-driven spiking neural networks and federated learning frameworks to effectively execute intelligent analysis tasks over amounts of distributed low-power devices but also perform vulnerability to poisoning attacks. The threat of backdoor attacks on traditional deep neural networks typically comes from time-invariant data. However, in FedNL, unknown threats may be hidden in time-varying spike signals. In this paper, we start to explore a novel vulnerability of FedNL-based systems with the concept of time division multiplexing, termed Spikewhisper, which allows attackers to evade detection as much as possible, as multiple malicious clients can imperceptibly poison with different triggers at different timeslices. In particular, the stealthiness of Spikewhisper is derived from the time-domain divisibility of global triggers, in which each malicious client pastes only one local trigger to a certain timeslice in the neuromorphic sample, and also the polarity and motion of each local trigger can be configured by attackers. Extensive experiments based on two different neuromorphic datasets demonstrate that the attack success rate of Spikewispher is higher than the temporally centralized attacks. Besides, it is validated that the effect of Spikewispher is sensitive to the trigger duration.
\end{abstract}
\begin{figure}
  \centering
\includegraphics[width=\textwidth]{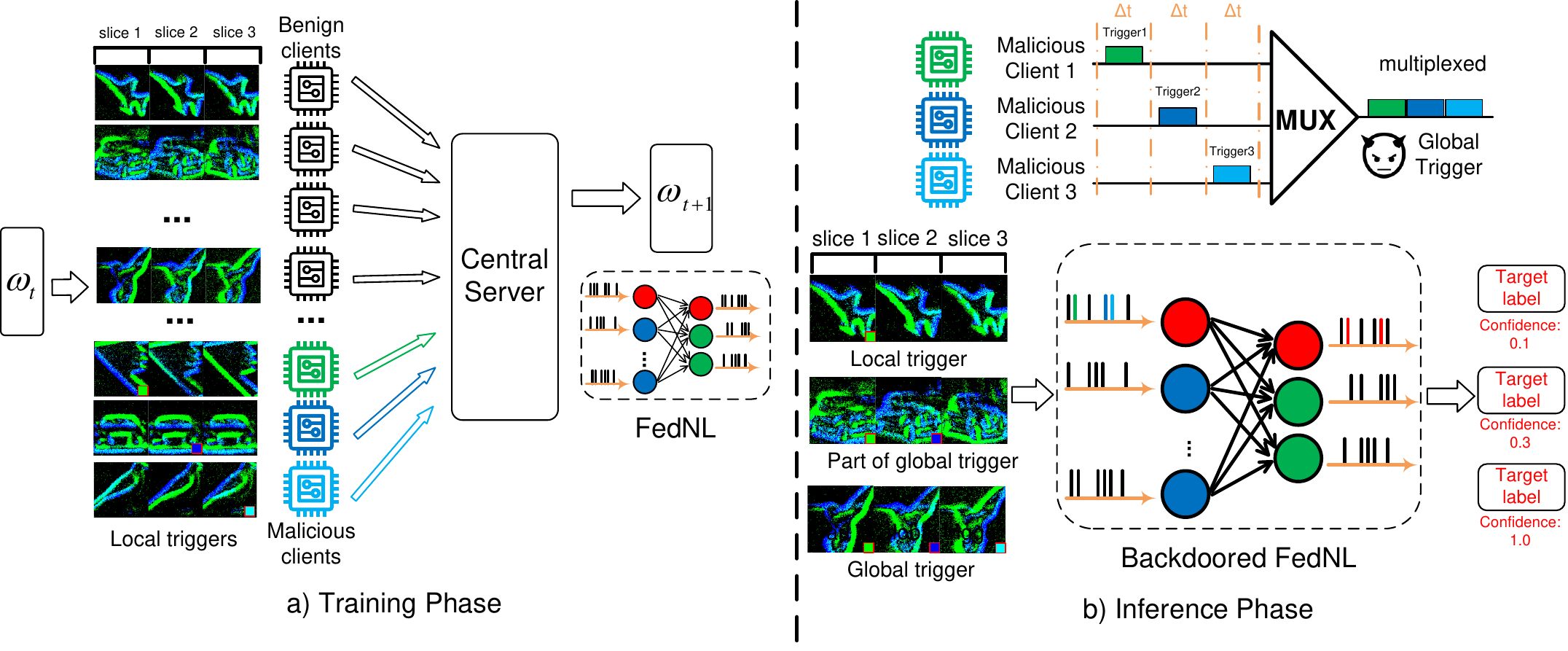}
  \caption{Overview of Temporal Spike Backdoor Attacks on FedNL. 
In the training phase, the Central Server aggregates parameters from local benign and malicious participants in the previous round $t$, updating the global SNN model parameters $\omega_{t+1}$. The attackers use only a subset of the global trigger's temporal sequence as the local trigger for implementing the backdoor attack. In the inference phase, all clients with the global SNN model will misclassify input with the global trigger into the target class.}
  \label{fig:1}
\end{figure}

\section{Introduction}
Federated neuromorphic learning (FedNL), a combination of federated learning (FL) and spiking neural networks (SNNs), enables substantial low-power devices to quickly and energy-efficiently obtain Artificial Intelligence (AI) models from large amounts of distributed data while protecting data privacy. In federated learning, each device trains an AI model locally and then uploads the model parameters or gradients to a central server for global model aggregation. SNNs have been proposed and explored as a low-power neuromorphic alternative to traditional deep neural networks (DNNs) due to the event-driven and discrete features of signal processing \cite{fang2021incorporating,eshraghian2023training}. In the era of thriving large models, for example, training the GPT-3 model consumed about 190,000 kWh of electricity \cite{dhar2020carbon}, FedNL emerges with the potential for allowing low-power devices to collaboratively train the large model.

Nowadays, FedNL has garnered widespread attention \cite{skatchkovsky2020federated,venkatesha2021federated,yang2022lead}. However, in the same way as the federated learning with DNNs \cite{melis2019exploiting,bagdasaryan2020backdoor}, federated neuromorphic learning is vulnerable to a variety of security threats, and one of the most critical threats is backdoor attack, which modifies the training set to inject triggers in certain examples. After training, the model will perform correctly in the main task. However, with the presence of triggers (backdoors) on the input samples, the model will go wrong and misclassify samples to the target label. Existing backdoor research has mainly considered DNNs rather than SNNs. Backdoor threats under FedNL urgently need further study.

Considerable attention has been paid to exploring backdoor attacks on traditional DNNs. However, there has been scant attention paid to investigating backdoor attacks targeting FedNL with SNNs.
In this paper, we delve into the feasibility of backdoor attacks in FedNL. In response to the temporally distributed characteristic, a novel method with the concept of Time Division Multiplexing for backdoor attacks is designed. By splitting the global trigger into multiple spike timeslices, the local triggers are concealed within each spike timeslice of the neuromorphic data, significantly enhancing the stealthiness and effectiveness of the backdoor attack. It exposes a novel security vulnerability for FedNL, which is crucial for the security of edge intelligence applications \cite{ansari2020security}.
The main contributions of this paper
are summarized as follows:
\begin{itemize}
\item A novel temporal spike backdoor attack scheme is proposed for FedNL over low-power devices, named Spikewhisper, which is distributed in the time dimension rather than the spatial dimension. To the best of our knowledge, this is the first work on the robustness and security of federated neuromorphic learning.
\item Different from traditional neural backdoor attacks, we identify that the backdoor effect of Spikewhisper is extremely sensitive to not only local trigger size and location but also temporal duration.
\item Extensive experiments on Attack Success Rate (ASR) and Main Task Accuracy (MTA) with two different neuromorphic datasets demonstrate that Spikewhisper achieves state-of-the-art attack effects against temporal centralized backdoor attacks.
\end{itemize}

The rest of this paper is organized as follows. Section 2 introduces the related work of FedNL and backdoor attacks. Section 3 introduces the Spikewhisper system model. The experiments are presented in Section 4. Section 5 provides a conclusion and outlook for our work.

\section{Related work}
Since our work primarily follows two research directions: Federated Neuromorphic Learning and Backdoor Attacks, a comprehensive introduction of recent advances in these two areas is as follows.
\subsection{Federated Neuromorphic Learning}
Federated Neuromorphic Learning (FedNL) is pioneered by Skatchkovsky et al. \cite{skatchkovsky2020federated} to effectively train SNNs for low-power edge intelligence, providing an effective trade-off between communication overhead and training accuracy. To capture dynamic spike characteristics at time-domain and reduce the training cost, Venkatesha et al.\cite{kim2021revisiting,venkatesha2021federated} proposed a Batch Normalization Through Time (BNTT) algorithm, which decoupled the parameters of each layer of neurons on the time axis. On the basics of this, they also validated that the accuracy of FedNL is 15\% higher than that of DNNs on CIFAR10 as well as the energy efficiency is 5.3 times higher. 
Xie et al. \cite{xie2022efficient} applied the FedSNN-NRFE approach based on neuronal receptive field encoding in traffic sign recognition. In comparison to CNN, FedSNN-NRFE significantly reduced energy consumption. Meanwhile, Yang et al. \cite{yang2022lead} proposed a lead federated neuromorphic learning framework for wireless edge intelligence, designating devices with high computational capacity, communication, and energy resources as leaders to effectively accelerate the training process. Wang et al. \cite{WANG2023126686} introduced SNNs into asynchronous federated learning, which adapts to the statistical heterogeneity of users and complex communication environments. 

\subsection{Backdoor Attacks}
Gu et al. first introduced neural backdoor attacks, named Badnets \cite{gu2019badnets}. This attack method involves training on images with square patches, creating a backdoor that can be triggered at will by the attacker. Bagdasaryan et al. introduced backdoor attacks into the field of FL \cite{bagdasaryan2020backdoor}. In this context, attackers train the backdoor model locally and upload local model updates scaled by a constraint replacing the benign global model. Bhagoji et al. proposed a stealthy model poisoning method \cite{bhagoji2019analyzing}, with the use of
an alternating minimization strategy that alternately optimizes for stealth and the adversarial objective. 
Xie et al. introduced the distributed backdoor attack (DBA) \cite{xie2020dba}, wherein the global trigger is spatially decomposed into local triggers. These local triggers are then individually embedded into the training data of multiple malicious parties, enabling a distributed implementation of the backdoor attack.

Abad et al. first investigated the application of backdoor attacks in SNNs using neuromorphic datasets \cite{abad2022poster,abad2024sneaky}. 
The subsequent work \cite{abad2024time} by Abad et al. concurrently explored backdoor attacks on FedNL alongside our research. Compared to this work, we made more explorations in the design and duration of local triggers.

In this paper, we identify a very sophisticated attack path with the concept of Time Division Multiplexing, that is, poisoning by different clients on different frames of neuromorphic data with different local triggers, which further improves the awareness level about the security risks of FedNL.

\section{System Model}
In this section, we will present the comprehensive system model of Spikewhisper. Section \ref{seciton3.1} introduces the backdoor threat model under FedNL, while Section \ref{section3.2} discusses the variations that attackers face when transitioning from FL to FedNL. Section \ref{section3.3} provides an overview of Spikewhisper.
\subsection{Threat Model}
\label{seciton3.1}
FedNL's goal is to train a global SNN model that can generalize well on test data $D_{test}$ after aggregating over the distributed training results from $N$ clients with their $N$ local datasets $D_{i}$ on a central server $S$. The objective of FedNL can be cast as a finite-sum optimization as below:
\begin{equation}
{\min_{\omega \in R^d} \left[ F(\omega) := \frac{1}{N}\sum_{i=1}^Nf_i(\omega) \right] }
\end{equation}
where $\omega$ stands for the parameters of the model and $f_i$ stands for the loss function $\sum_{(x,y)\in \mathcal{D}_i}\mathcal{L}(f_\omega(x), y)$.

Specifically, at round t, $S$ dispatches the current global SNN model $G_t$ to a subset of clients denoted by $n \in \left\{ 1, 2, ..., N \right\}$. The selected client, indexed as $i$, locally computes the loss function $f_i$ and adopts surrogate gradient descent \cite{neftci2019surrogate} for $E$ local epochs. 

\subsubsection{Attacker ability}
We consider the attacker's ability as follows:
\begin{itemize}
\item The attacker has full control over the local training data of any compromised participant.
All compromised participants conspire under the attacker's control to conduct backdoor attacks against the FedNL system, which is consistent with FL settings. \cite{wang2020attack}
\item The attacker can manipulate the local training process, such as updating hyperparameters like the number of epochs and learning rate.
\item The attacker can not tamper with any aspects of the benign participants' training.
\item The attacker does not have control over the central server's aggregation algorithm used to combine participants' updates into the joint SNN model.
\end{itemize}

\subsubsection{Attack Objective}
The attacker wants FedNL to produce a joint backdoor SNN model that has good performance on both the main task and the backdoor task. In other words, the SNN model predicts normally on any clean input while predicting a target label $\hat{y}$ on any input that has a global trigger. The adversarial objective for attacker $i$ in round $t$ with local dataset $D_i$ and target label $\hat{y}$ is:
\begin{equation}
\omega_{t+1} = \underset{\omega}{\arg\min} (\sum_{j\in D_i^{cln}} \mathcal{L}(f_\omega(x_j),y_j)  + \sum_{j\in D_i^{poi}} \mathcal{L}(f_\omega(\hat{x_j})),y_t))
\label{eq4}
\end{equation}

where the clean dataset $D_i^{cln}$ and poisoned dataset $D_i^{poi}$ satisfy $D_i^{cln} \cup D_i^{poi} = D_i$ and $D_i^{cln} \cap D_i^{poi} = \emptyset$. The $\hat{x}$ is the backdoored, and $y_t$ is the backdoor target label.

\subsection{Changes Faced by Attackers}
\label{section3.2}
The transition from FL to FedNL introduces a series of changes that pose challenges for attackers. These changes are outlined as follows.

The biggest difference between SNNs and traditional DNNs is the feature of information processing. DNNs process continuously changing real-value, whereas SNNs process discrete events that occur at certain time points due to using spiking neurons. The Leaky-Integrate-and-Fire (LIF) model \cite{gerstner2002spiking} is frequently used to simulate neuronal functions in SNNs.
\begin{figure}[h]
  \centering
\includegraphics[height=0.2\linewidth]{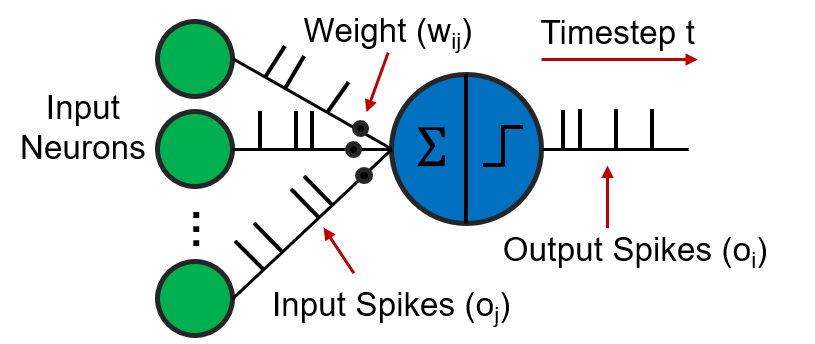}
  \caption{The Leaky-Integrate-and-Fire Behavior of Spiking Neuron $i$.}
  \label{fig:6}
\end{figure}

As shown in Fig. \ref{fig:6}, for given Spiking Neuron $i$, a set of spikes from $N$ input neurons is accumulated through the weight $w_{ij}$ for all $j\in N$, forming the membrane potential of the neuron. Once the membrane potential reaches the threshold $v$, an output spike is triggered. After the occurrence of a spike, the membrane potential is reset to the resting potential $u_{rest}$, or in the case of a soft reset, the membrane potential is decreased by the threshold $v$. The entire process persists for $T$ timesteps.

occurrence of a spike, the membrane potential is reset to the resting potential $u_{rest}$, or in the case of a soft reset, the membrane potential 
is decreased by the threshold $v$. The entire process persists for $T$ timesteps.
Formally, the Leaky Integrate-and-Fire (LIF) mechanism can be expressed as follows:

\begin{equation}
u_{i}^{t}=\lambda u_{i}^{t-1}+\sum_{j \in N} w_{i j} o_{j}^{t-1}
\label{eq2}
\end{equation}

where $u_i^t$ represents the membrane potential of neuron $i$ at timestep $t$, $\lambda$ is a constant leakage factor, indicating how much the membrane potential decreases per timestep. The discrete nature of information processing makes SNNs employ Surrogate Gradient Descent \cite{neftci2019surrogate} for training.

Neuromorphic data are widely considered to be the most suitable data for SNNs today. Neuromorphic data consists of many spiking events that are captured by the Dynamic Vision Sensor (DVS) sensing the intensity change (increase or decrease) of each pixel in the environment, e.g. $ON$ channel indicates an increase and $OFF$ channel indicates a decrease. The entire spiking sequence can be represented as an event of size $T \times P \times H \times W$, where $H$ and $W$ are the height and width, $T$ denotes the length of the recording time, and $P$ denotes two channels of polarity.
For ordinary data in the image domain, the triggers commonly are encoded in 256 possibilities per pixel per channel (usually 3 channels), which allows for many color combinations. In neuromorphic data, however, each pixel could only take the value 0 or 1 in two channels (On and OFF polarity channel). In other words, the trigger space is reduced to 4 possibilities encoded by the two different polarities in SNNs, which limits the backdoor trigger design greatly. Additionally, neuromorphic data is time-encoded, which is temporal and contains $T$ frames while ordinary images are static and non-temporal.

After being aware of the discrete nature of information processing, the limitations of color combinations, and the flexibility in the time dimension, we propose Spikewhisper to inject backdoors in FedNL.
\subsection{Spikewhisper Framework}
\label{section3.3}


Our proposed Spikewhisper utilizes the concept of Time Division Multiplexing (TDM) to enhance the backdoor efficacy and stealthiness in FedNL as shown in Fig. \ref{fig:1}. In TDM, different signals are interleaved within different time slots to share a channel. Similarly, in Spikewhisper, the neuromorphic data is segmented into multiple timeslices. Let $T$ represent the total duration of the neuromorphic data. The allocation of timeslices can be represented as follows: 
\begin{equation}
T = \{t_1,t_2,...,t_K\} \text{ while } T=\sum_{i=1}^K \Delta t_i
\label{eq12}
\end{equation}
Where $\Delta t_i$ represents the duration of timeslice $t_i$ and $K$ represents the total number of timeslices, consistent with the number of malicious participants controlled by the attacker. The length of timeslice allocated to each malicious participant can be freely distributed according to demand, constrained only by the total length $T$.
In the field of communications, there is the concept of channel utilization rate. In Spikewhisper, the redundant space for backdoor triggers of neuromorphic data can be analogized as a 'temporal channel', and the temporal utilization rate $U$, can be represented as:
\begin{equation}
U=\sum_{i=1}^K \frac{\Delta L_i}{\Delta t_i}
\label{eq11}
\end{equation}
Where $\Delta L_i$ stands for the local trigger's duration of $t_i$. Subsequent experiments in Section \ref{4.5.1} illustrate that high utilization produces a stronger backdoor effect.


In the poisoned data design stage, the attacker has the flexibility to configure the polarity and motion of the local triggers in a wide range of cases. As stated in Section \ref{section3.2}, 
We denote these four polarity possibilities as $p_0$, $p_1$, $p_2$, and $p_3$. Consequently, for different polarity combinations, triggers show different colors, i.e. black, green, dark blue, or light blue. Due to the multiple time frames in neuromorphic data, we can change the position of local triggers frame by frame, creating a sense of motion. This aligns more with the motion (illumination changes) nature of neuromorphic data and facilitates more stealthy and more natural triggers. Formally, the process of generating a poisoned data is as follows:
\begin{equation}
(\hat{x},y_t) = R_i((x,y),p,m,s)
\label{eq9}
\end{equation}
Where $\hat{x}$ represents the modified input, $y_t$ is the target label, $p$ denotes the trigger polarity, $m$ represents the trigger's motion trajectory, and $s$ represents the size of the trigger.

For the backdoor training phase, the malicious client adopts the batch poison approach. Based on the poisoning rate $r$, the malicious client $i$ poisons $r \times BatchSize$ clean data, only poisoning the data at the i-th timeslice. Additionally, the malicious client uses a smaller learning rate and more local epochs to achieve a better backdoor effect. Formally, the training process is represented as follows:
\begin{equation}
\Delta w_{i j}=\sum_{t=1}^{T} \frac{\partial L}{\partial w_{i j}^{t}}=\sum_{t=1}^{T} \frac{\partial L}{\partial o_{i}^{t}} \frac{\partial o_{i}^{t}}{\partial u_{i}^{t}} \frac{\partial u_{i}^{t}}{\partial w_{i j}^{t}}
\label{eq10}
\end{equation}
Where $\Delta w_{i j}$, the gradient of the weight connecting neuron $i$ and $j$ is accumulated over $T$ timesteps. The loss function $L$ is to evaluate the mean square error between output fire rates and sample label $y_i$, which is given by
\begin{equation}
L=\frac{1}{N}\sum_{i=1}^N\left(\frac1T\sum_{t=1}^TS_i(t)-y_i\right)^2
\label{eq8}
\end{equation}

In the inference phase of FedNL, the attacker uses a global trigger to implement a backdoor attack. The global trigger is formed by aggregating all local triggers, which can be represented as $T_{\mathrm{global}}=\sum_{i=1}^{K}T_{\mathrm{local},i}$

\begin{figure}[!h]
  \centering
    \includegraphics[width=0.7\linewidth]{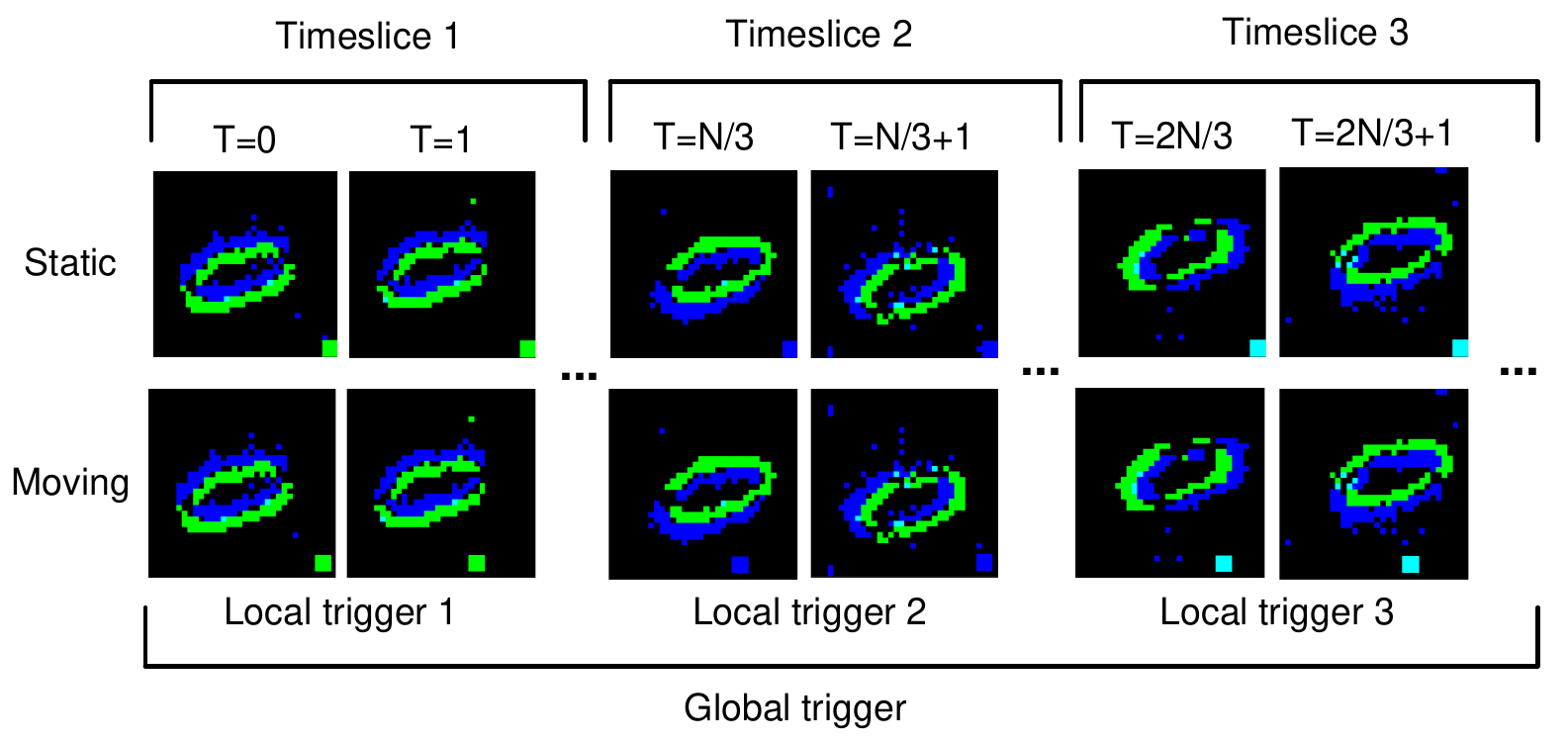}
  \caption{Static trigger and moving trigger used in Spikewhisper.}
  \label{fig:2}
\end{figure}

In the Spikewhisper framework, the allocation of timeslices and the design of triggers exhibit a high degree of flexibility. For ease of experimental demonstration, we adopt equally sized time slice allocations.
Specifically, we segmented neuromorphic data into 
3 timeslices and designed two types of global triggers, namely static trigger and moving trigger, as illustrated in Fig. \ref{fig:2}.
Taking the static trigger as an example, this trigger is inspired by BadNets \cite{gu2019badnets} in DNNs, where a fixed-position square trigger is on all frames of neuromorphic data. However, the polarity of the trigger varies over timeslices. As the polarity changes from $p_1$ to $p_3$ (except the polarity $p_0$ representing the color black), the color of the static trigger changes on different timeslices too.
The moving trigger is inspired by the previous backdoor work on SNNs \cite{abad2022poster,abad2024sneaky}.
The trigger horizontally and smoothly changes the location from frame to frame, creating an effect of moving among the actions of the image, making the procedure more stealthy and more natural. As for the variation in the polarity for each timeslice of moving triggers, it remains consistent with that of static triggers, which is different from the previous SNN backdoors.

If we transplant the traditional FL backdoor attack to FedNL, we would obtain a form of temporally centralized attack. In this scenario, each malicious user employs the same global trigger to poison the data, and this trigger spans every frame of the neuromorphic data. In contrast to Spikewhisper, which conceals local triggers within specific data timeslices, this attack method significantly increases the risk of poisoning exposure, exhibits poor stealthiness, and overlooks the temporal distribution characteristics of FedNL. 

\section{Experiment}
\subsection{Datasets \& Network Architectures}

We evaluate Spikewhisper on two neuromorphic datasets: N-MNIST \cite{orchard2015converting} and CIFAR10-DVS \cite{li2017cifar10}, which are converted from the most popular benchmarking datasets for AI security in computer vision. 
\begin{figure}[htb]
    \centering
    
    \subcaptionbox{N-MNIST\protect\label{fig:7a}}
      {\includegraphics[width=0.45\linewidth]{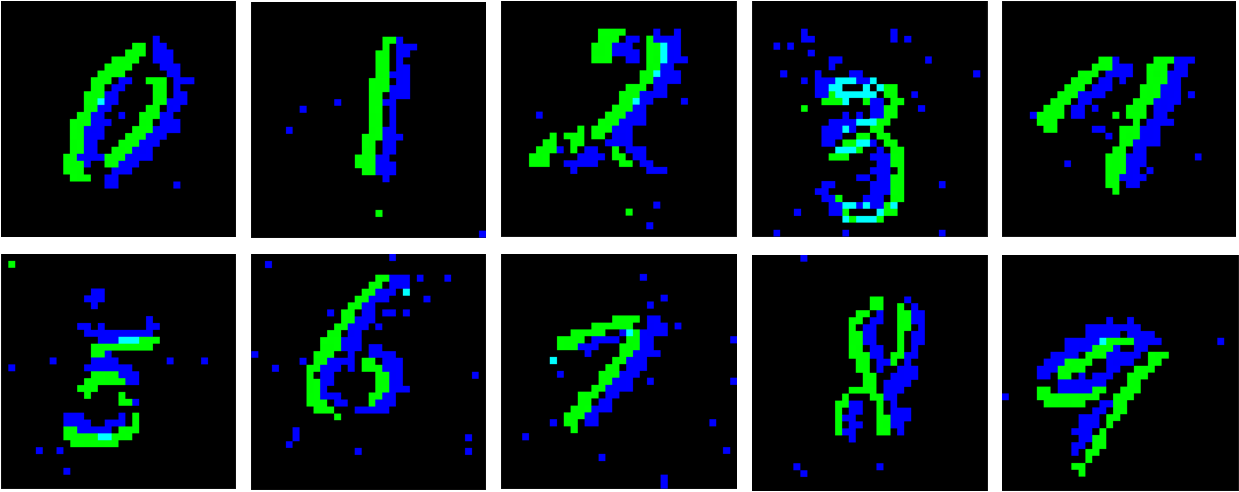}}
    \subcaptionbox{CIFAR10-DVS\protect\label{fig:7b}}{\includegraphics[width=0.45\linewidth]{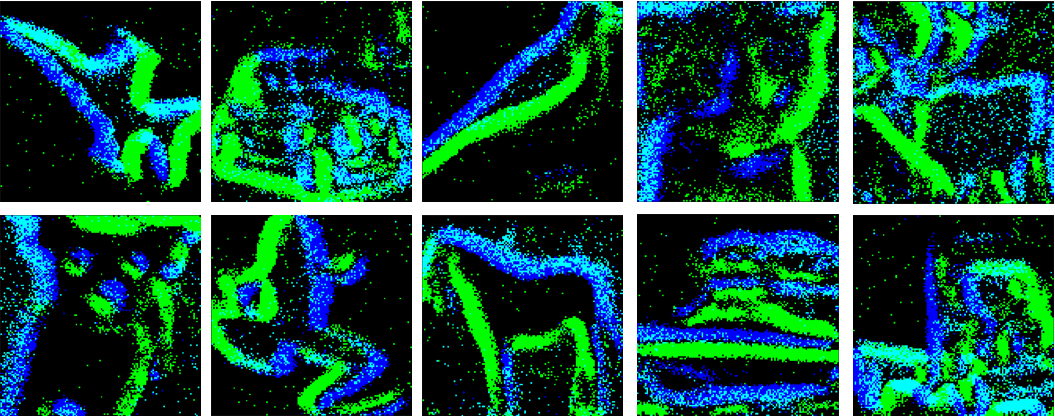}}
    \label{fig:7}
    \caption{Neuromorphic Data Samples (one frame of each sample).}
\end{figure}

The N-MNIST dataset is a spiking version of the MNIST \cite{lecun1998mnist} dataset. It comprises $60000$ training samples and $10000$ test samples for 10 classes, maintaining the same scale as the original MNIST dataset, but with a sample size of $34\times34$ instead of $28\times28$. Likewise, the CIFAR10-DVS dataset is a spiking version of the CIFAR10 \cite{krizhevsky2009learning} dataset. It contains $10000$ $128\times128$ samples, and $1000$ samples per class, corresponding to 10 classes. The sample size for both the N-MNIST and CIFAR10-DVS datasets is represented as $T \times P \times H \times W$, where $T$ is the time steps (we set $T = 18$ in the experiments), $P$ is the polarity, $H$ is the height, and $W$ is the width.

We employed distinct network architectures for the two datasets. For the N-MNIST dataset, the network comprises two convolutional layers followed by batch normalization and max pooling layers, then two linear layers with dropout, concluding with a voting layer aimed at enhancing classification robustness. In the case of the CIFAR10-DVS dataset, the network consists of four convolutional layers incorporating batch normalization and max pooling layers, along with two linear layers incorporating dropout, and similarly concludes with a voting layer.

\subsection{Experiment Setup}

We used SpikingJelly \cite{fang2023spikingjelly} framework to implement the SNN model and partition the N-MNIST and CIFAR10-DVS datasets into T=18 frames. Within FedNL, we employed the Adam optimizer with a local learning rate $l_r$ and batch size of $B$ for training over $E$ local epochs. 

Following the multiple-shot attack setup of Bagdasaryan et al. \cite{bagdasaryan2020backdoor}, 
the attackers need to undergo multiple rounds of selection, and the accumulation of malicious updates is necessary for the success of the attack. Otherwise, the backdoor will be weakened by benign updates and quickly forgotten by the global SNN model. To expedite the convergence speed of backdoor learning and quickly observe the distinctions between temporally centralized attacks and Spikewhisper, we consistently select attackers in each training round. Then we randomly choose benign participants to form a total of 10 participants.
Furthermore, we expedite the training speed by employing IID distribution to allocate the neuromorphic datasets among a total of 50 participants. 
To ensure training stability, malicious participants engage in backdoor training after a certain number of benign training rounds (10 for N-MNIST, and 25 for CIFAR10-DVS). 


In our experiments, we employ the same 
static and moving global triggers to assess the attack success rates of Spikewhisper and 
temporally centralized attacks. To ensure a fair comparison, we make certain that the total number of backdoor trigger pixels for Spikewhisper attackers is identical to that of temporally centralized attackers. 

\subsection{Evaluation Metrics}
We evaluate Spikewhisper and Temporally Centralized Attacks with the commonly used metrics:
\begin{itemize}
\item \textbf{Attack Success Rate (ASR)} represents the percentage of attacked samples that the compromised model successfully predicts as the desired target label.
\item 
\textbf{Main Task Accuracy} signifies the precision with which the infected model predicts benign test samples.
\end{itemize}
\subsection{Experiment Result}

\begin{figure}[tb]
    \centering
    \subcaptionbox{Static Trigger\protect\label{fig:3a}}
      {\includegraphics[width=0.4\linewidth]{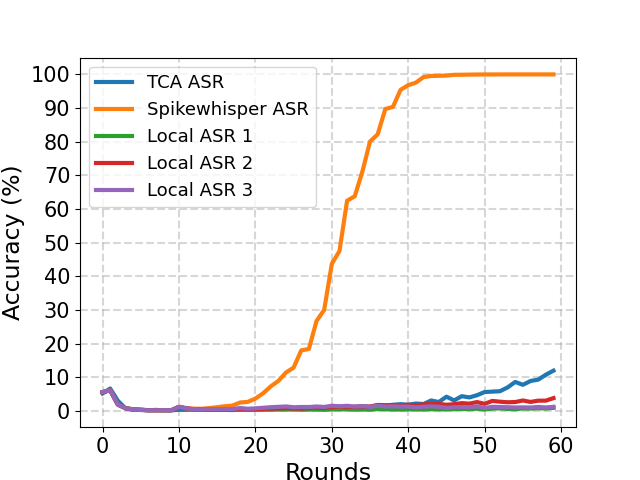}}
    \subcaptionbox{Moving Trigger\protect\label{fig:3b}}
      {\includegraphics[width=0.4\linewidth]{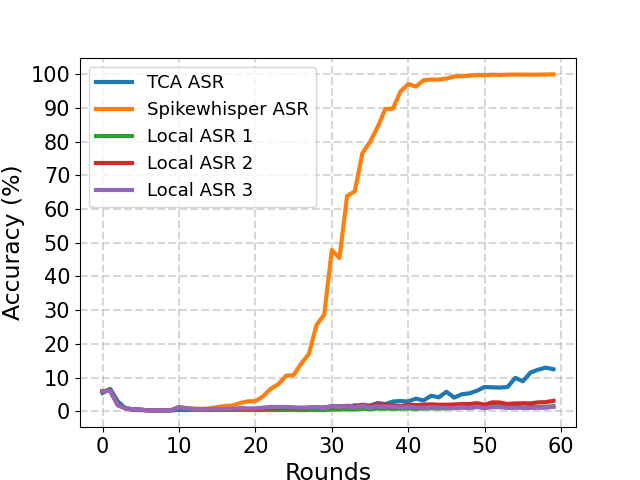}}
    \caption{Spikewhisper and Temporally Centralized Attacks (TCA) on 
the N-MNIST dataset}
    \label{fig:3}
\end{figure}

\begin{figure*}[!tb]
    \centering
    \subcaptionbox{Static Trigger\protect\label{fig:4a}}
      {\includegraphics[width=0.4\linewidth]{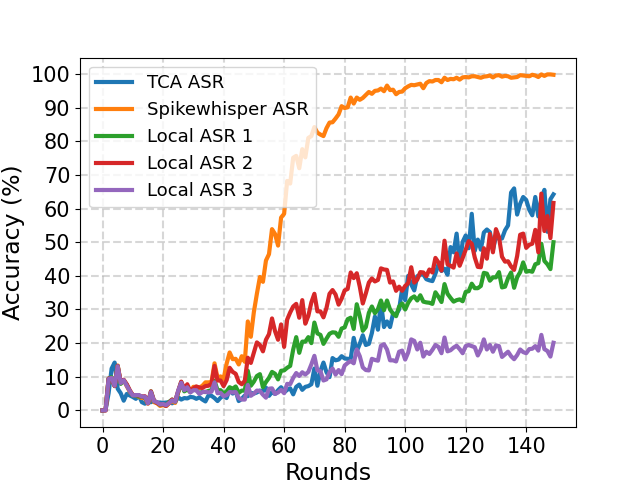}}
    \subcaptionbox{Moving Trigger\protect\label{fig:4b}}
      {\includegraphics[width=0.4\linewidth]{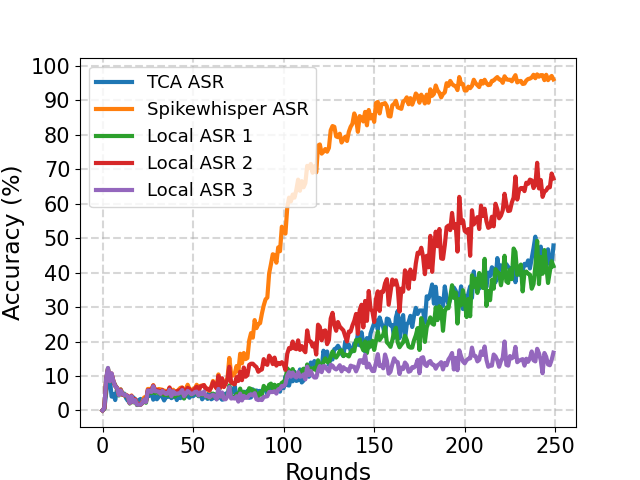}}
    \caption{Spikewhisper and Temporally Centralized Attacks (TCA) on 
the CIFAR10-DVS dataset}
    \label{fig:4}
\end{figure*}

The experiment conducted involved attacks using both static and moving triggers on the N-MNIST and CIFAR10-DVS datasets, evaluating the impact on federated neuromorphic learning through multiple rounds of SNN model aggregation. The results demonstrate the efficacy of Spikewhisper and the comparative performance of Spikewhisper versus the centralized approach.




As shown in Fig. \ref{fig:3}, for the N-MNIST dataset, Spikewhisper with static trigger exhibited a rapid escalation in the attack success rate (ASR) for the global trigger, surpassing 99\% by the 42nd round. In contrast, the local static triggers showed significantly lower ASRs at 0.49\%, 1.99\%, and 1.27\%, respectively. The centralized static attack yielded a relatively low ASR of 2.09\% at this point, highlighting the inferiority compared to Spikewhisper. Spikewhisper with moving trigger on N-MNIST also achieved high ASRs for the global trigger, exceeding 99\% by the 46th round. The centralized attack achieved an ASR of only 12.46\% in the final round, indicating a failure to establish a backdoor in the global SNN model, while Spikewhisper reached 100\% ASR.

Switching to the CIFAR10-DVS dataset as shown in Fig. \ref{fig:4}, Spikewhisper with static trigger also demonstrated a rapid rise in the ASR of the global trigger, exceeding 99\% by the 120th round, with local static triggers exhibiting varying 
low ASRs. The temporally centralized attack lagging behind at 50\% ASR at this point. The moving trigger attacks on CIFAR10-DVS were more challenging, with a slower growth rate in ASR compared to static triggers. 
In Spikewhisper, the global moving trigger reached 95\% ASR by the 190th round, with the ASR for the temporally centralized attack less than 40\%. Even in the final round, the temporally centralized attack achieved an ASR of only 47.89\%, emphasizing the effectiveness of Spikewhisper with its higher ASR.

\begin{table}[h]
\centering
\caption{Main Task Accuracy (\%)}
\label{tab:3}

\resizebox{0.85\columnwidth}{!}{%
\begin{tabular}{ccc}
\hline

Attack Types                         & N-MNIST & CIFAR10-DVS                                                                 \\ \hline
Baseline (No Attack)                             & 98.96   & \begin{tabular}[c]{@{}c@{}}64.30 (150 Round)\\ 65.60 (250 Round)\end{tabular} \\ 
Spikewhisper w/ Static Trigger           & 98.87   & 62.20 (150 Round)                                                            \\ 
Temporally Centralized Attack w/ Static Trigger & 98.99   & 62.80 (150 Round)                                                            \\ 
Spikewhisper w/ Moving Trigger           & 98.76   & 64.30 (250 Round)                                                            \\ 
Temporally Centralized Attack w/ Moving Trigger & 98.92   & 65.50 (250 Round)                                                            \\ \hline
\end{tabular}%
}
\end{table}

As for the impact of Spikewhisper and temporally centralized attacks on main task accuracy, please refer to Table \ref{tab:3}. From the table, we can observe that both Spikewhisper and temporally centralized attacks have almost negligible effects on the accuracy of the main task. The decrease in main task accuracy caused by Spikewhisper does not exceed 2.1\%, while the accuracy under temporally centralized attacks at times, even surpasses the baseline due to fluctuation.

In summary, Spikewhisper successfully injected a backdoor into the global SNN model, demonstrating its potential threat to federated neuromorphic learning systems. By implementing Spikewhisper on the N-MNIST and CIFAR10-DVS datasets, we achieved attack success rates (ASRs) of over 99\%, with little impact on the accuracy of the main task. Compared to temporally centralized attacks, the ASR of Spikewhisper was significantly higher in all experimental instances. In the N-MNIST dataset, even with local trigger attack success rates consistently below 10\%, the global trigger achieved a 99\% ASR. This indicates that temporally centralized attacks are inefficient in federated neuromorphic learning.

\subsection{Ablation Study}
In this subsection, we investigated the impact of the temporal length of Spikewhisper triggers on the backdoor effects based on ASRs. Additionally, we experimented with the attack performance of Spikewhisper in the Non-IID setting, further exploring the generality of Spikewhisper.

\subsubsection{Temporal Utilization of Trigger}
\label{4.5.1}
In the preceding experiment setup, each data sample consists of $T=18$ frames. The global trigger also spans $18$ frames, distributing to three local triggers, each lasting for $6$ frames, achieving a 100\% temporal utilization rate.

\begin{figure*}[!bthp]
    \centering
    \subcaptionbox{Static Trigger on N-MNIST\protect\label{fig:5a}}
      {\includegraphics[width=0.24\linewidth]{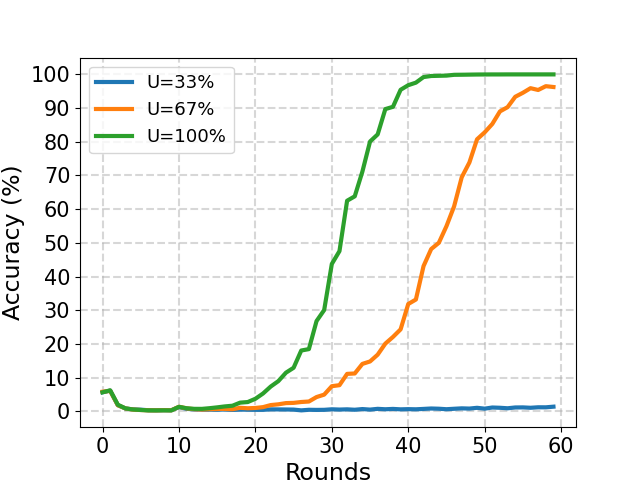}}
    \subcaptionbox{Moving Trigger on N-MNIST\protect\label{fig:5b}}
      {\includegraphics[width=0.24\linewidth]{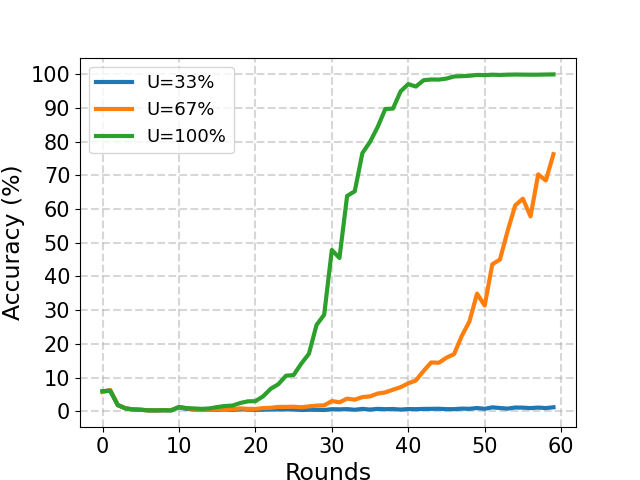}}
    \subcaptionbox{Static Trigger on CIFAR10-DVS\protect\label{fig:5c}}
      {\includegraphics[width=0.24\linewidth]{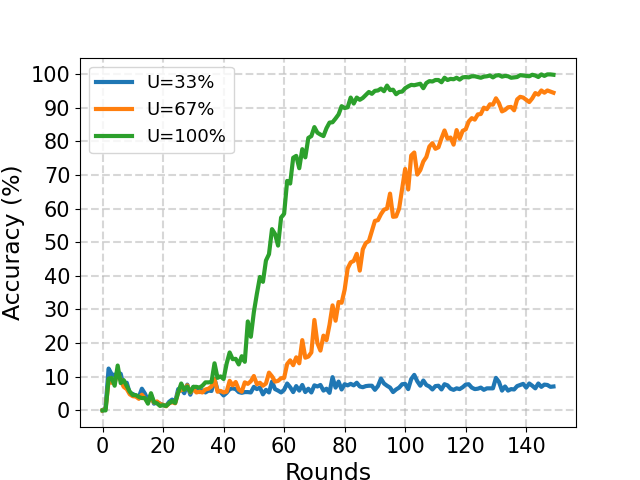}}
      \subcaptionbox{Moving Trigger on CIFAR10-DVS\protect\label{fig:5d}}
      {\includegraphics[width=0.24\linewidth]{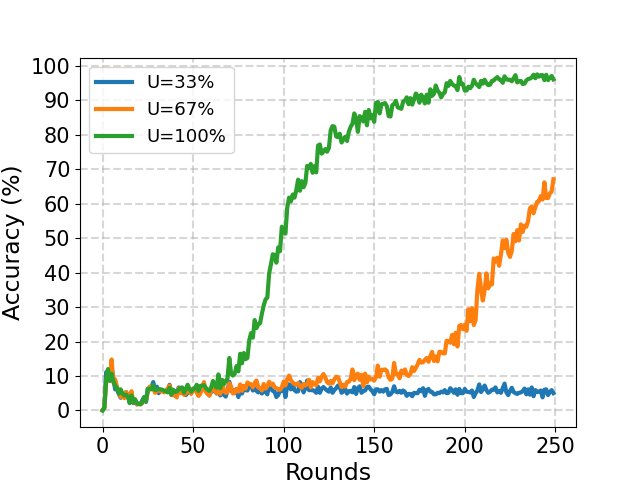}}
    \caption{Effects of the temporal utilization rate $U$ on Attack Success Rate of Spikewhisper. $U=0\%$ means no data poisoning, and $U=100\%$ means each local trigger persists throughout the entire $6$ frames, collectively forming an $18$ frames global trigger.}
    \label{fig:5}
\end{figure*}

 We maintain the presence of three malicious participants, inserting triggers in the first, middle, and last thirds of the data frames (each comprising $6$ frames). $U=0\%$ indicates no insertion of local triggers, meaning no data poisoning. U=33\% and U=67\% respectively represent inserting $2$ frames and $4$ frames of local triggers per timeslice, and $U=100\%$ signifies that each local trigger persists throughout the entire $6$ frames, collectively forming an $18$ frames global trigger.

The experimental results of two triggers on two neuromorphic datasets are shown in Fig. \ref{fig:5}. It can be observed that all four plots exhibit similar phenomena, where the backdoor effect of Spikewhisper gradually strengthens with the increase of $U$. When $U=33\%$, both trigger types on the two datasets fail to inject backdoors into the global SNN model, resulting in minimal backdoor effects. When $U=67\%$, successful backdoor effects can be achieved, but more rounds are required to fully inject the backdoor into the global SNN model. When $U=100\%$, the injection speed of the backdoor reaches its maximum, and the number of rounds required for backdoor learning convergence is also minimized.

Based on the experimental analysis of the impact of $U$ on the backdoor effect, it can be concluded that the longer the duration of the trigger, the more likely the success of the attack in the federated neuromorphic learning scenario. This success leads to the injection of a backdoor into the global SNN model.

\subsubsection{Trigger Size and Location}
In backdoor attacks on DNNs, the size and location of the trigger can significantly impact the attack effectiveness. To identify the difference of Spikewhisper against these backdoors in DNNs, in Section \ref{4.5.1}, we observe that Spikewhisper is only sensitive to the temporal duration of the trigger, which is a unique feature under FedNL. In this section, we will explore whether the size and location of the trigger similarly affect the effectiveness of Spikewhisper.

\begin{table}[h]
\caption{ASR of Spikewhisper with different trigger sizes and locations on N-MNIST.}
\label{tab:4}
\centering
\resizebox{0.6\columnwidth}{!}{%
\begin{tabular}{cccc}
\hline
Trigger size & Position     & Type   & ASR (60 Rounds) \\ \hline
$1\times1$          & bottom-right & static & 0.28\%          \\
$1\times1$          & bottom-right & moving & 0.90\%          \\
$2\times2$          & bottom-right & static & 56.44\%         \\
$2\times2$          & bottom-right & moving & 87.96\%         \\
$3\times3$          & bottom-right & static & 100.00\%        \\
$3\times3$          & bottom-right & moving & 100.00\%        \\
$3\times3$          & middle       & static & 0.26\%          \\
$3\times3$          & middle       & moving & 0.26\%          \\
$3\times3$          & top-left     & static & 100.00\%        \\
$3\times3$          & top-left     & moving & 99.91\%        \\ \hline
\end{tabular}%
}
\end{table}

Under different trigger sizes and position configurations, the ASR of Spikewhisper on the N-MNIST is presented in Table \ref{tab:4}. 
Under 60 rounds, it can be observed that the size and position of triggers significantly influence the backdoor performance of Spikewhisper. In terms of size, the ASR increases sequentially as the trigger size grows. Interestingly, in previous experiments, we observed that the backdoor convergence speed with moving triggers was slower than that of static triggers. However, in the case of smaller triggers, we found that moving triggers could induce a stronger backdoor effect. Regarding position, we moved the trigger from its original position in the bottom right corner of the data along the diagonal. We tested three positions: bottom-right, middle, and top-left. It can be seen that triggers at corners exhibit excellent ASR, while those in the middle perform poorly. We attribute this to the fact that the main body of N-MNIST samples is located in the middle of the images, affecting the effectiveness of triggers in the middle.

\subsubsection{Non-IID Scenario}

In real-world application scenarios of federated learning, the data distribution among each participant is often non-independent and identically distributed (Non-IID) \cite{li2022federated}. Therefore, we also evaluate the Spikewhisper's performance under Non-IID settings to validate its practicality.

\begin{figure}[htb]
    \centering
    \subcaptionbox{Static Trigger\protect\label{fig:8a}}
      {\includegraphics[width=0.4\linewidth]{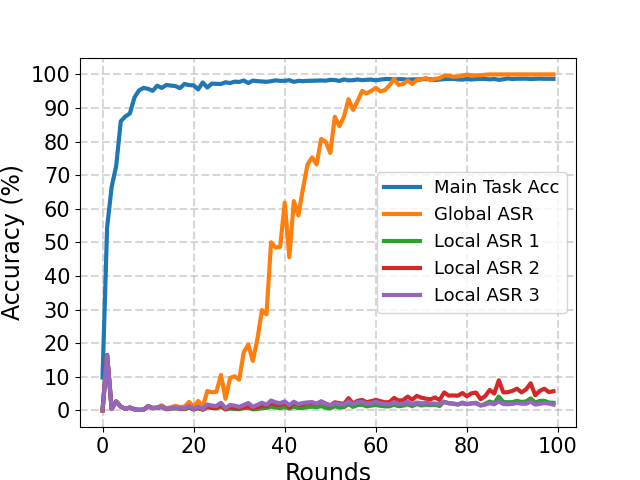}}
    \subcaptionbox{Moving Trigger\protect\label{fig:8b}}{\includegraphics[width=0.4\linewidth]{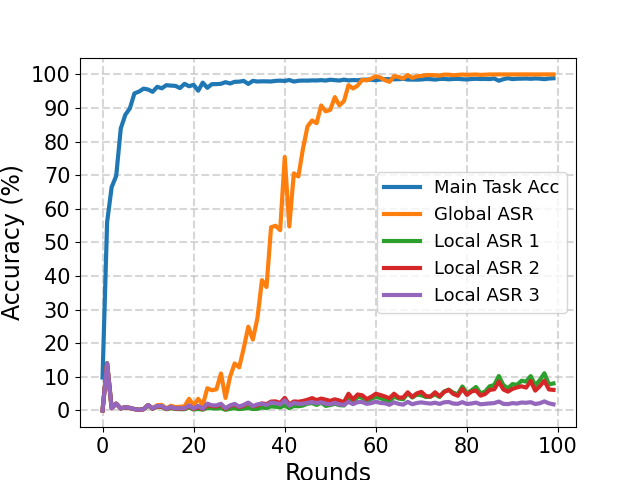}}
    \label{fig:8}
    \caption{Spikewhisper on N-MNIST in Non-IID Scenario.}
\end{figure}

To evaluate the performance of Spikewhisper in the Non-IID scenario, we utilized a Dirichlet distribution \cite{minka2000estimating} with $\alpha=0.5$ hyperparameter to partition the 60,000 training samples of the N-MNIST dataset. 

It can be observed that in the Non-IID setting, Spikewhisper successfully injected a backdoor into the global SNN model, whether using the static trigger or the moving trigger. In comparison to the IID setting, the communication rounds required for the ASR to reach 99\% increased from around 40 rounds to approximately 60 rounds. However, the consistent phenomenon remains: when Spikewhisper successfully injects the backdoor, the ASR for each local trigger remains at an extremely low level.

\section{Conclusions \& Future Work}


This paper aims to investigate a novel temporal spike backdoor attack named Spikewhisper in FedNL over low-power devices, particularly utilizing the distributed nature of federated learning and the temporal characteristics of spiking neural networks with the concept of time division multiplexing. We evaluate Spikewhisper using static and moving triggers on two different neuromorphic datasets: 1) N-MNIST and 2) CIFAR10-DVS. The results indicate that Spikewhisper outperforms temporally centralized attacks significantly, achieving an attack success rate of over 99\%. We study the temporal duration of triggers in Spikewhisper, revealing that the more frames the triggers occupy, the stronger the resulting backdoor effect, facilitating faster injection of backdoors into the global SNN model. Furthermore, we explore the impact of trigger size and location on Spikewhisper. Lastly, we also evaluate the performance of Spikewhisper in the Non-IID setting. Our research indicates that federated neuromorphic Learning is susceptible to temporal spike backdoor attacks (Spikewhisper), yet there is currently a lack of dedicated backdoor defense measures in this domain. The development of corresponding defense strategies specifically targeting spiking neural networks and neuromorphic data in FedNL is poised to become a crucial direction for future research.

\bibliographystyle{splncs04}
\bibliography{mybibliography}





\end{document}